# Impact of Plasmonic Modes and Thermophysical Properties on the Double-Pulse Structuring of Highly-Ordered LIPSS for Biosensing Applications

G.D. Tsibidis[a]*, F. Fraggelakis[a], P. Lingos[a], E. Cusworth[b], V.G. Kravets[b], A.N. Grigorenko[b], A.V. Kabashin[c]*, E. Stratakis[a,d]*

[a] Institute of Electronic Structure and Laser (IESL), Foundation for Research and Technology (FORTH), Heraklion, Crete 70013, Greece; [b] Department of Physics and Astronomy, Manchester University, Manchester M13 9PL, U.K.; [c] Aix Marseille Univ, CNRS, LP3, Marseille, 13288, France; [d] Department of Physics, University of Crete, Heraklion, Crete 71003, Greece

* E-mail: tsibidis@iesl.forth.gr; andrei.kabashin@univ-amu.fr; stratak@iesl.forth.gr

## ABSTRACT

The fabrication of highly ordered laser-induced periodic surface structures (LIPSS) on thin metallic films is dictated, predomaninatly, by a synergy of periodic electromagnetic energy deposition and complex fluid dynamics. In this work, we present a combined experimental and theoretical study on the formation of ultra-regular LIPSS on 32-nm-thick Au films using a double-pulse femtosecond laser scheme. We demonstrate that for thin films, the excitation of coupled Surface Plasmon Polaritons (SPPs) at both interfaces dictates the initial energy distribution. On the othe hand, the final morphology is greatly influenced by hydrodynamical processes. Interestingly, due to the low electron-phonon coupling of Au and the high thermal confinement of the thin film, single-pulse irradiation leads to uncontrolled hydrodynamic instabilities a non uniform topographies. Thus, we demonstrate that a double-pulse approach with an optimized interpulse delay ($\Delta\tau \approx 1.2$ ns) effectively controls the melt duration and viscosity, suppressing complex fluid motion and promoting the growth of highly ordered arrays. These structures support narrow surface lattice resonances (SLRs) suitable for high-sensitivity plasmonic biosensing.

## 1. INTRODUCTION

Laser-induced periodic surface structures (LIPSS) have evolved from a phenomenon of fundamental interest in ultrafast laser-matter interactions into a well-established platform for nanoscale surface functionalization. While the formation of ripples on bulk metallic targets is generally understood as an interference process between the incident laser field and surface electromagnetic waves [1], the extension of this technique to ultra-thin metallic films (i.e. with thickness comparable to the optical penetration depth [2, 3]) introduces a significantly more complex physical landscape. Thin films are of paramount importance for the development of integrated plasmonic devices and lab-on-a-chip biosensors [4, 5]; however, their reduced dimensionality, fundamentally, alters both the electromagnetic energy deposition and the subsequent thermophysical evolution of the material [2, 6, 7].

In principle, the investigation of the transition from a bulk substrate to a thin-film geometry necessitates a re-evaluation of the plasmonic modes which are involved in patterning. In a thin-film system, more specifically, the air/gold/glass architecture investigated in this paper, the electromagnetic response is no longer dictated solely by the air/metal interface. By contrast, as the film thickness approaches the skin depth of the metal, the evanescent fields at the top and bottom interfaces couple, giving rise to symmetric and anti-symmetric Surface Plasmon Polariton (SPP) modes [2, 3]. These coupled modes not only shift the resonant periodicity of the resulting LIPSS but also concentrate the absorbed energy within a confined volume, intensifying the thermal load compared to bulk gold [7]. This spatial confinement enables the creation of highly sensitive plasmonic transducers, it also drives the material into a regime of extreme hydrodynamic instability. On the other hand, the challenge of nanostructuring gold (Au) is further influenced by its intrinsic thermophysical properties. Gold is characterized by an exceptionally low electron-phonon coupling constant [6, 8] and a very high electron thermal conductivity [6, 8]. Simulations presented in a previous report manifest a deeper modulation of the height of LIPSS in the thin-film case as a result of the combined effect of the geometric confinement and gold's thermophysical properties upon irradiation of the solid with ultrashort laser pulses (Figure 1) [2].

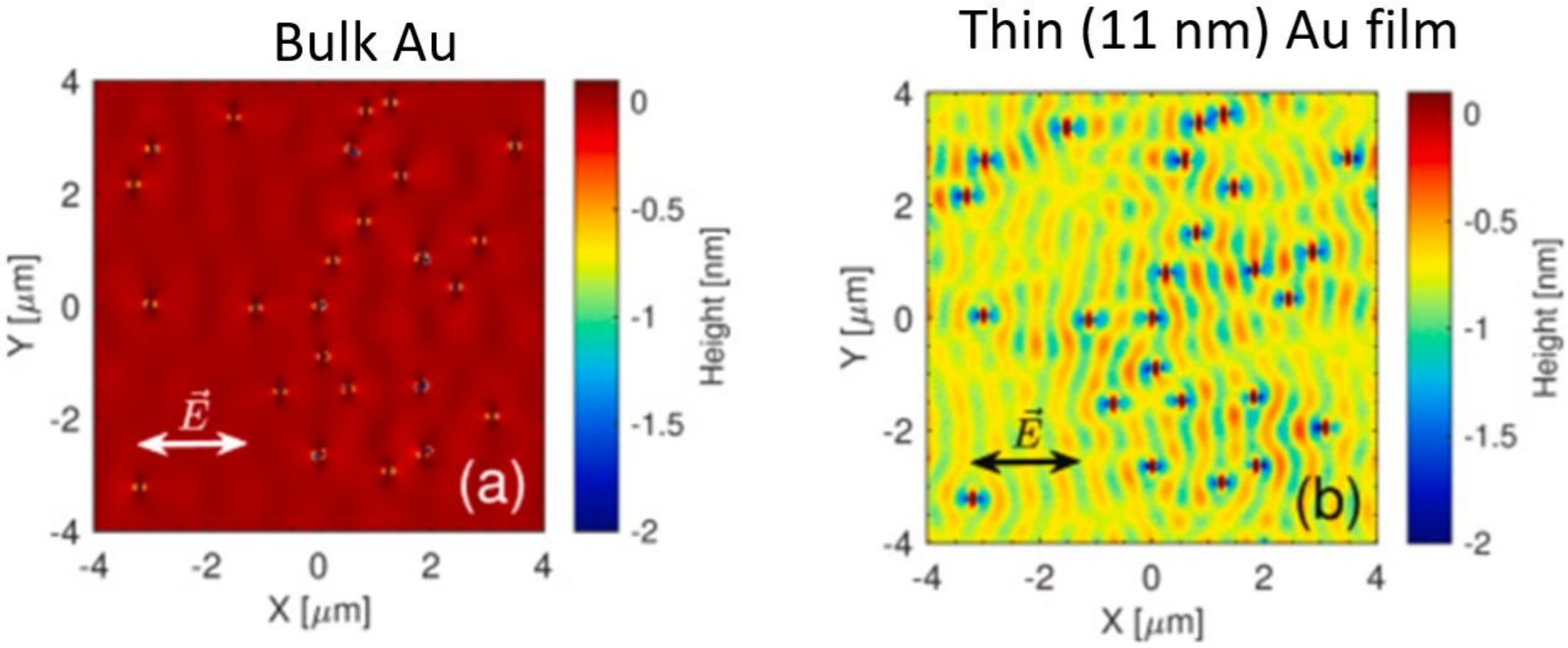

Figure 1: LIPSS patterns for (a) bulk, (b) thin (thickness $d$ = 11 nm) Au film. Double-headed arrow indicates the orientation of the laser beam polarization.

Furthermore, upon femtosecond excitation, the electron subsystem reaches high temperatures while the lattice remains relatively cold for several picoseconds. This 'bottleneck' in energy transfer, combined with the thermal insulation provided by a glass substrate leads to a prolonged molten state. Once the lattice temperature exceeds the melting point, the thin liquid layer becomes susceptible to strong hydrodynamic forces. The most prominent physical process characterizing, then, the response of the material is related to the produced Marangoni effect [9, 10]. Because surface tension in liquid gold is strongly temperature-dependent, the periodic temperature gradients-initially established by plasmonic interference induce a lateral flow of melt from hot regions toward cooler ones. As highlighted by the simulated results (Figure 1), the synergistic effects of hot spot formation, localized energy concentration and complex hydrodynamic interactions drive the evolution of non-uniform surface topographies. These mechanisms not only constrain the reproducibility and uniformity of the induced pattern but also carry important implications for practical applications, in particularly in processes where controlled surface structuring, material stability, and functional performance are critical.

To overcome the above hydrodynamic limitations, temporal pulse shaping via a double-pulse irradiation (DPI) scheme is presented as a strategy to generate highly-ordered LIPSS on thin Au films. It is shown that by splitting the laser energy into two pulses separated by a pico (to nano) second-scale delay ($\Delta\tau$), it is possible to gain a degree of freedom to manipulate the material's transient thermodynamic state [11]. In this work, we present a comprehensive ivestigation that combines the experimental fabrication of highly-ordered LIPSS on thin Au films with a multiscale theoretical framework. We emphasize the synergy between the excitation of coupled SPPs and the control of the dynamics of the produced liquid phase in regulating the induced pattern uniformity. By tailoring the fluid dynamics through double-pulse delays, we demonstrate the fabrication of long-range periodic arrays highly-ordered LIPSS on thin Au films that support ultra-narrow Surface Lattice Resonances (SLRs). These structures provide a high-throughput, lithography-free pathway toward the next generation of ultrasensitive plasmonic biosensors.

## 2. EXPERIMENTAL METHODS

### 2.1 Experimental Setup and Patterning Protocol

The nanostructuring of thin Au films was performed using a high-repetition-rate Yb:KGW femtosecond laser system (Pharos, Light Conversion), delivering pulses at a central wavelength of $\lambda_L$=1030 nm with a pulse duration of $\tau_p$=170 fs and a repetition rate of 5 kHz. To investigate the transition from irregular to ordered patterning, a double-pulse irradiation configuration was implemented using a Michelson-type interferometer. The incident beam was split into two pulses of equal energy (P1=P2). The interpulse delay, $\Delta\tau$, was precisely controlled via a motorized translation stage in one of the interferometer arms, allowing for temporal separations ranging from 0 (effectively single-pulse) to 2 ns. The target samples were 32-nm-thick Au films deposited on glass substrates. This specific thickness was chosen based on theoretical predictions to produce efficient coupling of surface plasmon polaritons at the Air/Au and Au/Glass interfaces [11]. The peak fluence was maintained between $\Phi$=0.08-0.32J/cm² to ensure that only the Au film was modified without damaging the underlying glass substrate.

### 2.2 SEM Analysis: Single vs. Double Pulse Irradiation

The surface morphology of the laser-processed areas was characterized using Scanning Electron Microscopy (SEM). The experimental results reveal a dramatic dependence of the structural quality on the temporal pulse configuration [11].

**Single Pulse Structuring (SPS) Results:**

As shown in the SEM imagages for $\Delta\tau$=0 (Fig. 2), single-pulse irradiation results in highly irregular and periodic structures with a number of 'hot spots'. While a dominant periodicity is observed, the patterned topography is, also, characterized by the presence of dark islands where ablation has occurred. The high-resolution SEM images show that the Au film has undergone a process of thermocapillary flow, where the continuous liquid layer breaks into isolated droplets along the ridges of the EM-induced interference pattern. This morphology is indicative of violent fluid dynamics: the extreme temperature gradients produced by the localized energy deposition trigger ablation and strong thermocapillary flows. These flows drive the molten Au from the peaks of the intensity distribution toward the valleys with sufficient momentum to cause ablation (see theoretical investigation in the next Section).

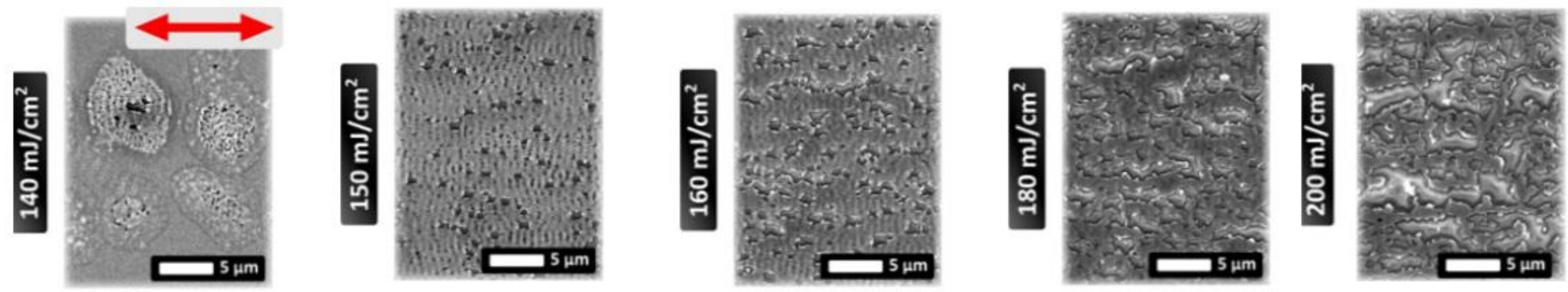

Figure 2: LIPSS patterns at different fluences assuming single pulses. Double-headed arrow indicates the orientation of the laser beam polarization.

**Double Pulse Structuring (DPS) Results:**

By contrast, the introduction of a delay $\Delta\tau$ (Fig. 3) produces highly-regular, continuous LIPSS for some combinations of the laser parameters (i.e. special choice of pulse separation and fluence). The SEM analysis for topographies obtained in those conditions shows that the irregular ripple segments in SPI are replaced by long-range highly-uniform periodic patterns that extend across the entire laser-scanned area. This suggests that the double-pulse protocol successfully manages the fluid dynamic evolution of the melt. By spreading the energy absorption over a longer timescale, the peak Marangoni stress is reduced, allowing the molten Au to redistribute into a periodic relief without reaching the thermocapillary flow threshold.

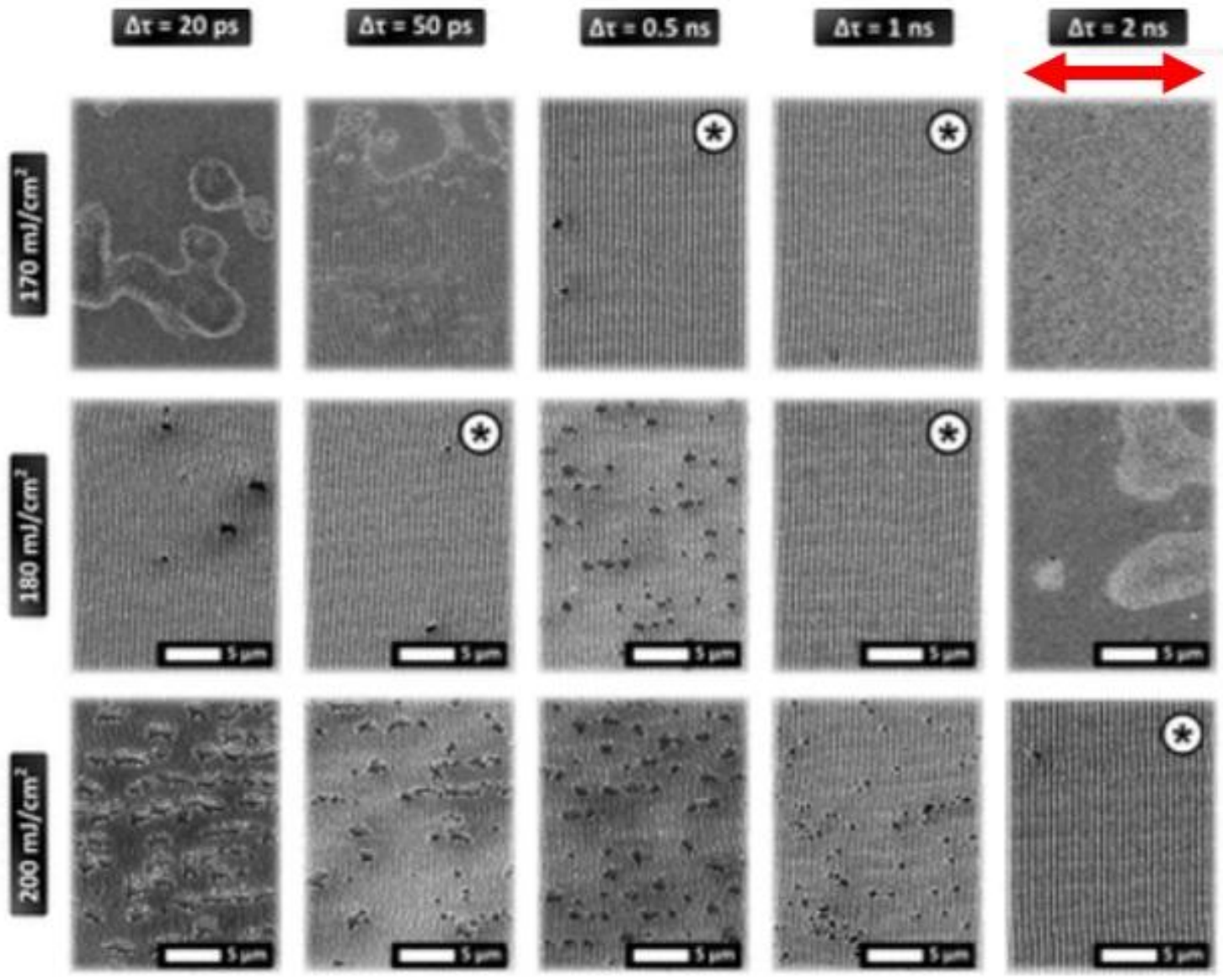

Figure 3: LIPSS patterns at different fluences and pulse separations. Double-headed arrow indicates the orientation of the laser beam polarization. Homogenous LIPSS patterns are produced for conditions marked with a star (*).

## 3. THEORETICAL FRAMEWORK: THE SYNERGY OF ELECTROMAGNETIC WAVES, THERMAL AND FLUID DYNAMICS

To explain the formation of highly-ordered LIPSS on thin Au films, a multiscale model was used to describe the dynamics from the initial electromagnetic absorption to the molten material transport and pattern formation. The process is governed by a three-step synergy: (i) coupled plasmonic energy deposition, (ii) electron excitation and electron-lattice thermalization and (iii) Marangoni flow and resolidification.

### 3.1 Electromagnetic Energy Deposition: Coupled Plasmonic Modes

In bulk metals, LIPSS periodicity is typically explained by the interference between the incident laser beam and SPPs excited at the air/metal interface. However, for a 32-nm thin film, the electromagnetic (EM) landscape is fundamentally altered. As shown in previous studies [2], the proximity of the metal/substrate interface allows for the excitation of coupled SPP modes. The spatial distribution of the electromagnetic field $\vec{\mathbf{E}}(\vec{\mathbf{r}},\mathbf{t})$ is calculated by solving Maxwell's equations using a 3D-FDTD approach [2, 5, 11, 12]. On the other hand, for a thin film, the dispersion relation for the coupled TM-polarized SPP modes is given by:

$$\exp(-2k_m d) = \frac{k_m/\varepsilon_m + k_a/\varepsilon_a}{k_m/\varepsilon_m - k_a/\varepsilon_a} \times \frac{k_m/\varepsilon_m + k_S/\varepsilon_S}{k_m/\varepsilon_m - k_S/\varepsilon_S} \quad (1)$$

where d is the thickness of the film, $k_j = \sqrt{\tilde{\beta}^2 - \varepsilon_j k_0^2}$ (j= 'a' 'm', 'S', for 'air', 'metal', 'substrate'), $\varepsilon_j$ stands for the dielectric permittivity of the j material, $k_0$ ($=2\pi/\lambda_L$) is the free-space wavenumber at the laser wavelength and $\tilde{\beta}$ is the propagation constant of the SP. The interference between the refracted beam and these coupled modes creates a periodic intensity distribution.

The numerical solution of Eq.1 provides the values of the supported SPP wavelength $\Lambda = 2\pi/\mathrm{Re}(\tilde{\beta})$ for a given film thickness. The calculated wavelengths of the two-interface surface plasmons are dependent on the film thickness, as illustrated in **Error! Reference source not found.**. Each interface can sustain bound SPPs and symmetric and antisymmetric modes are supported [13]. For $d = 32$ nm, two SPP wavelengths are predicted, 687 nm and 1026 nm, closely matching the periodicity of the simulated energy absorption patterns and experimentally measured values. Both theoretical predictions exhibit strong agreement with experimental measurements, thereby reinforcing the electromagnetic origin of the observed structures.

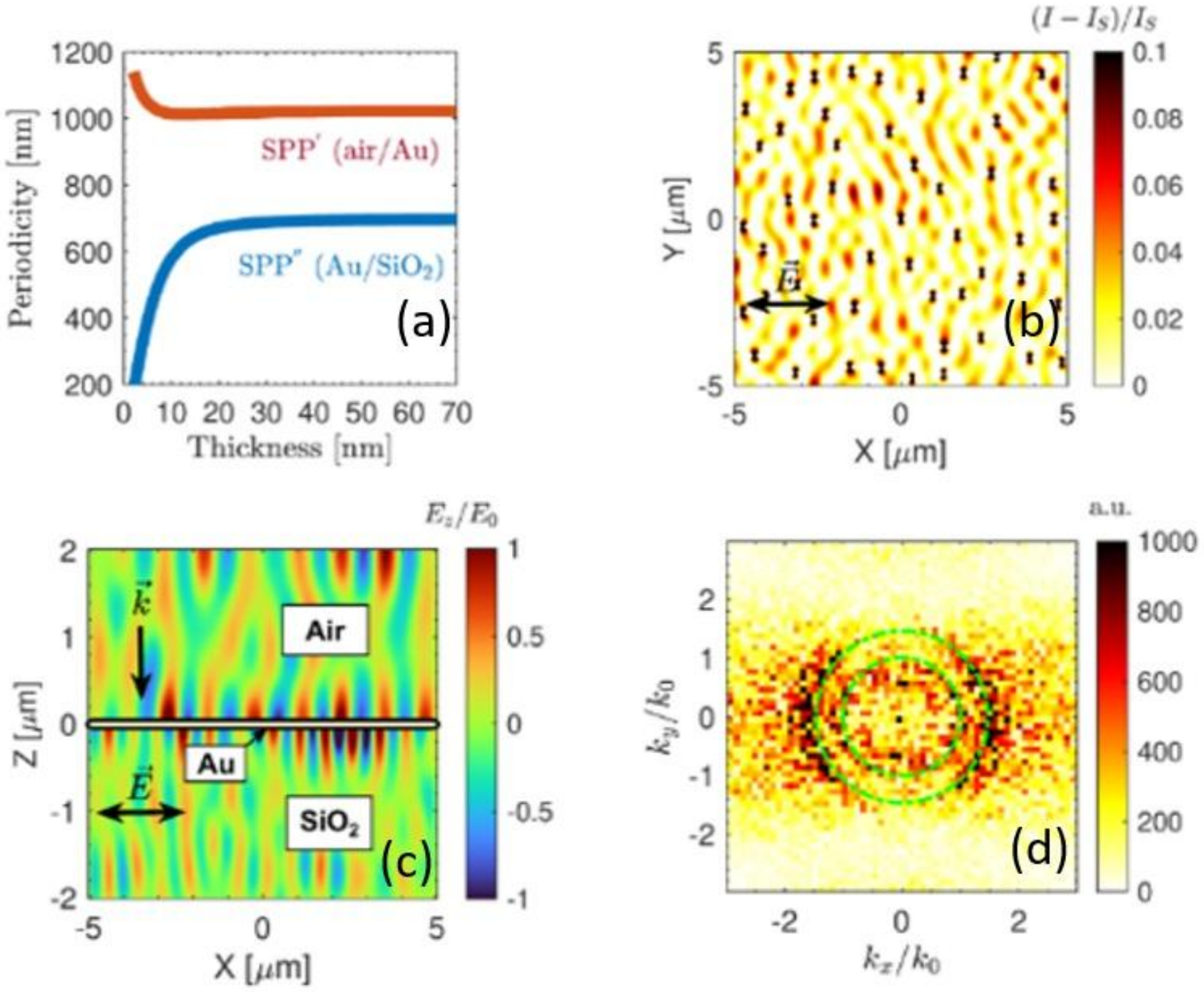

Figure 4: (a) Periodicities of SPP at air/Au and Au/$SiO_2$ interfaces as a function of the Au film thickness, (b) the electromagnetic profile and the produced near- and far-fields are illustrated in the transverse plane, (c) distribution of the electric field in air, metal and dielectric material on the propagation plane, (d) periodicities of the electromagnetic waves.

To investigate the characteristics of the excited surface waves which are the precursors to both the periodicity and orientation of LIPSS, an initially non-planar surface was considered. Surface roughness on Au was modeled as a random arrangement of semispherical protrusions with radius R=16 nm [2, 7, 11, 14]. Electromagnetic simulations indicate that the two distinct LIPSS periods arise from surface plasmon polariton modes excited at the Au-substrate and air-Au interfaces, respectively. More specifically, both near-field and far-field modes are generated. A rigorous evaluation of the electromagnetic response was performed using the Finite Integration Technique (FIT) to solve Maxwell's equations, implemented through the commercial software CST Studio Suite. In this study, nanobumps were selected as scattering centers instead of nanoholes. In ultra-thin films, nanoholes can induce direct coupling to the bottom interface, complicating the interpretation of interfacial plasmonic effects. Smaller holes would produce comparable field distributions but with a reduced SPP excitation efficiency due to weaker absorption, whereas larger nanobumps would promote multipolar localized plasmon modes. The resulting electromagnetic field profiles and corresponding near- and far-field distributions in the transverse plane are presented in Figure 4b, while the electric field distributions across air, metal, and dielectric regions along the propagation plane are shown in Figure 4c. A Fast Fourier Transform (FFT) analysis in $\mathrm{k_x}/\mathrm{k_0}(\approx \lambda_\mathrm{L}/\Lambda)$ space ($\lambda_\mathrm{L}, \Lambda$ correspond to the laser wavelength and the periodicity of the modes, respectively) reveals a dual-periodicity signature consistent with experimental observations (Figure 4d).

### 3.2 Thermal Dynamics, Marangoni Flow and Film Stability

To describe the multiscale processes leading to surface pattern formation upon irradiation of a two-layered Au film/substrate system with femtosecond laser pulses, a theoretical model is employed [6, 7, 8]. A detailed description of the model in case of a two-layered material comprising of a thin film of Au placed on $SiO_2$ has been presented in a recent report [11]. The inclusion of the pulse separation in the intensity profile is via the use of the expression $I_{total}(x,y,z,t) \sim \left[ I(x,y,z) e^{-4\log 2\left(\frac{t-3\tau_p}{\tau_p}\right)^2} + I'(x,y,z) e^{-4\log 2\left(\frac{t-3\tau_p-\Delta\tau}{\tau_p}\right)^2} \right]$. On the other hand, $I(x,y,z)$ and $I'(x,y,z)$ represent the spatial intensity distributions of each pulse, calculated from electromagnetic simulations and appropriately projected so that each pulse carries half of the total energy. For a single pulse, the total intensity reduces to: $I_{total}(x,y,z,t)$ is $\sim I(x,y,z) e^{-4\log 2\left(\frac{t-3\tau_p}{\tau_p}\right)^2}$. The subsequent excitation and thermal response of the $Au/SiO_2$ system under various irradiation conditions is modeled using the Two-Temperature Model (TTM) [11, 15]). It is noted that the fluence values used in the work can lead to either a phase transition (i.e. melting) or ablation (i.e. mass removal). It is noted that the onset of ablation considered in this work occurs when the lattice temperature exceeds $0.9T_c$=5625 K [6] where $T_c$ (=6250 K) stands for the critical point of Au.

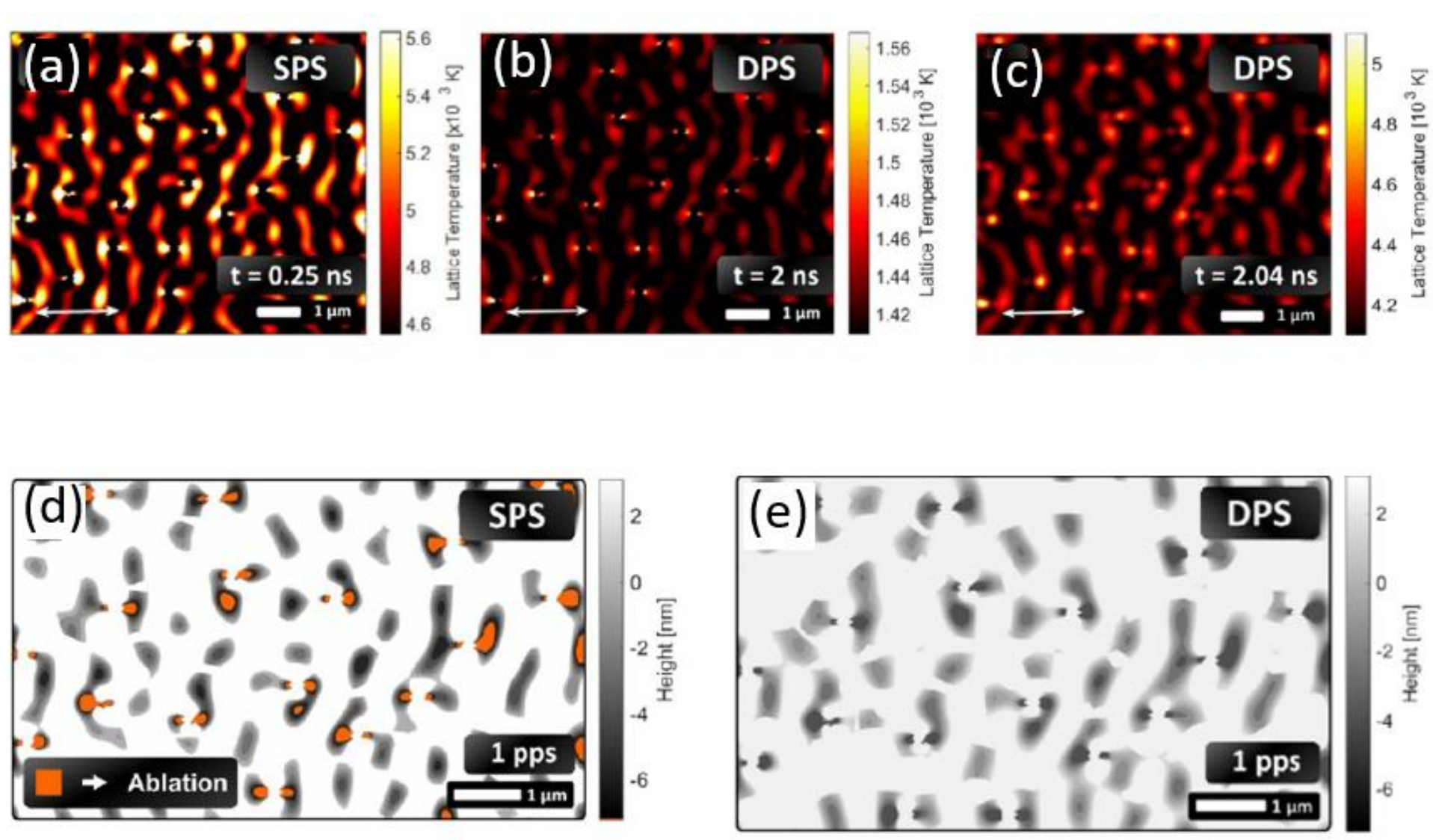

Figure 5: (a) Lattice temperature spatial profile on the surface of Au for SPS at t = 0.25 ns (white regions show ablated part). Lattice temperature spatial profile on the surface of Au at t = 2 ns (b) and t ~2.04 ns (c) (the white double-headed arrow indicates the laser polarization). Simulation results of surface relief resulting from irradiation with Φ = 280 mJ/cm$^2$ for a single pulse for SPS (d) and a single pair of pulses for DPS (e). Orange-colored areas indicate ablation in (d).

The surface modification is analyzed through the dynamics of hydrothermal waves generated by the produced phase transitions. The laser conditions used in the simulations $\Phi = 0.28$ J/cm$^2$ are appropriately selected to ensure the formation of a sufficiently large volume of molten Au after the exposure to the first pulse ($\Phi/2 = 0.14$ J/cm$^2$). The generation and evolution of hydrothermal waves and the dynamics of the produced fluid movement as a result for the phase transformation is mathematically described by the Navier-Stokes Equation (NSE)

$$\rho_0 \left( \frac{\partial \vec{u}}{\partial t} + \vec{u} \cdot \vec{\nabla} \vec{u} \right) = \vec{\nabla} \cdot \left( -P + \mu \left( \vec{\nabla} \vec{u} \right) + \mu \left( \vec{\nabla} \vec{u} \right)^T \right) \quad (2)$$

where $\mu$ and $\rho_0$ stand for the viscosity and density, respectively, of the molten (uncompressed) material, while P and $\vec{u}$ are the pressure and velocity of the fluid and superscript T denotes the transpose of the vector $\vec{\nabla}\vec{u}$. The solution of NSE is conducted through the employment of appropriate thermocapillary boundary conditions

$$\frac{\partial u}{\partial Z} = -\sigma/\mu \frac{\partial T_L}{\partial X} \text{ and } \frac{\partial v}{\partial Z} = -\sigma/\mu \frac{\partial T_L}{\partial Y} \quad (3)$$

at the liquid free surface where (u,v,w) are the components of $\vec{u}$ in Cartesian coordinates. In Eq.3, σ stands for the surface tension of the material. The inhomogeneous deposition of energy induces Marangoni effect redistributing the material from hotter to cooler regions. By employing Eq. 3, we obtain the lattice temperature profile on the Au surface as shown in Figures 5a-c. This thermal spatial distribution predicts the formation of a molten surface profile at t=2 ns resulting in a height pattern corresponding to the hydrothermal waves generated by single and double laser pulses as illustrated in Figures 5d–e.

In summary, the first pulse pre-heats the material and initiates a controlled melting phase while the second pulse arrives at a moment when the material's optical and hydrodynamic properties have been modified. This strategy effectively controls both the melting procedure and the dynamics of the molten volume that accounts for the generation of the LIPSS; thus, by optimizing the delay, we can suppress the peak temperature gradients that drive irregular Marangoni flows and therefore, an appropriate control of the melt front is feasible.

## 4. PLASMONIC PERFORMANCE: SURFACE LATTICE RESONANCES (SLR)

The ability to fabricate large-area, highly uniform periodic arrays offers a significant potential for the excitation of ultra-narrow Surface Lattice Resonances (SLRs). These resonances arise from the diffraction-mediated coupling of individual localized plasmon resonances (LPRs) across nanoparticles or nanostripes resulting in exceptional optical sensitivity. Such properties make SLRs particularly attractive for applications in biosensing and advanced photonic devices [11, 16]. More specifically, the resonances are observed when diffracted orders from the periodic lattice couple to the SPPs of individual gold ridges, producing ultra-narrow spectral features. In general, the diffraction-mediated coupling of on the order of one hundred resonators is adequate to produce relatively narrow surface lattice resonances, whereas extending the coupling to one thousand or more resonators can further compress the resonance linewidth to approximately 1-2 nm (FWHM) [11, 16]. In the present work, we investigated the optical response of laser-fabricated LIPSS to evaluate whether the structural uniformity and periodic order of the resulting arrays are sufficient to sustain high-quality SLRs. To this end, the ellipsometric parameters $\Psi$ and $\Delta$ were measured over a broad spectral range (550-1700 nm) as a function of the angle of incidence.

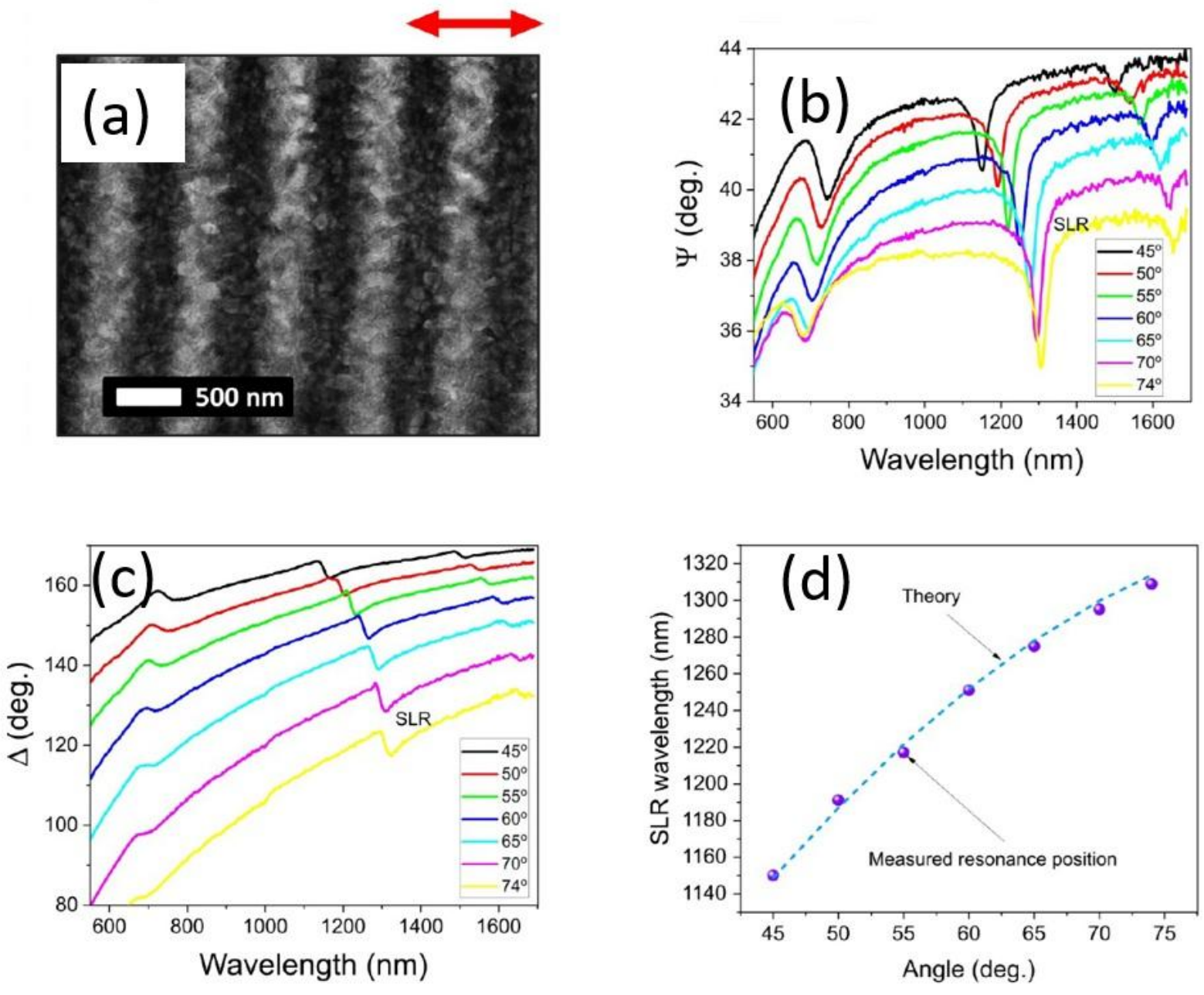

Figure 6: SEM (a) image of a sample prepared at a fluence of Φ = 170 mJ/cm$^2$ and time delay Δτ = 500 ps, (b) Ellipsometric parameter Ψ (amplitude) as a function of the incident wavelength and angle for the sample shown in panel (a). SLR denotes the positions of the plasmonic surface lattice resonances, (c) Ellipsometric parameter Δ (phase) as a function of the incident wavelength and angle for the same sample, (d) Comparison of the measured resonance position of SLR (magenta circles) and theory (dotted line).

In our tests, we used an optimized structure produced by DPS at a fluence of $\Phi$=170 mJ/cm$^2$ and time delay $\Delta\tau$ = 500 ps (Figure 6a) where the produced periodicity was measured to be equal to 655±20 nm. Figure 6b,c show the ellipsometry parameters measured on this sample. Three sets of collective resonances are observed at wavelengths of ~700, ~1250 and ~1600 nm. While the relatively weak resonances around 700 and 1600 nm can be attributed to coupling of incident light to SPPs on gold, a pronounced resonance around 1250 nm is obviously due to diffraction coupling of localized plasmons excited in the fabricated periodic nanostructure. Indeed, these essentially plasmonic SLRs have a spectral position described by the formula $\lambda_{Res} = \Lambda(n + \sin\theta)$ (Figure 6d), where $\Lambda$ is the period of the structure, n is the effective refractive index for the light beam diffracted along the surface of the sample, and $\theta$ is the angle of incidence.

According to results illustrated in Figure 6b, the observed SLR signatures exhibit a pronounced spectral sharpness with a full width at half-maximum of approximately 20 nm. This corresponds to a quality factor of Q=65 [11] which exceeds significantly that of the uncoupled LPR mode, whose linewidth typically ranges between 80 and 100 nm (corresponding to Q~10). The remarkably reduced linewidth of the SLR demonstrates the effectiveness of collective diffractive coupling within the array. Such spectrally narrow resonances, when supported by well-ordered metamaterial lattices can enhance sensing performance by at least an order of magnitude compared to uncoupled LPRs in disordered nanoparticle ensembles [17]. This significant improvement emphasises the potential of the observed SLRs for high-sensitivity biosensing applications.

On the other hand, Figure 6c indicates that phase jumps occur at the minima of the SLRs although their magnitude remains modest (below ~8°) due to the incomplete suppression of reflected intensity at resonance. Since sharper phase discontinuities require deeper intensity minima, the relatively shallow resonances observed here naturally limit the phase response. This behavior is linked to the geometry of the laser-structured features whose dimensions (~32 nm) are constrained by the initial Au film thickness and are smaller than the optimal 100-130 nm (i.e. typically required for 'tolopogical darkness'). Further optimization of femtosecond laser processing could enable structuring of thicker films and larger plasmonic elements (100-120 nm), allowing the combination of ultranarrow SLRs with strong phase singularities. Such architectures may also support topologically dark quasi-resonant modes with dramatically enhanced refractive index sensitivity, making their fabrication a promising direction for future development.

## 5. DISCUSSION

The evolution from disordered to highly-uniform LIPSS represents a fundamental advancement in the strategy for ultrafast surface engineering on thin films. This work demonstrates that the final surface morphology is not dictated by a single physical mechanism but rather by a precise synergy between electromagnetic energy deposition, thermophysical relaxation and fluid dynamic mass transport. Theoretical simulations demonstrated that the thin-film configuration dictates the excitation of coupled SPP modes at both the air/Au and Au/glass interfaces. On the other hand, the experimental results manifest that the response to the EM alone is insufficient for high-quality patterning. The 'bottleneck' in this process is the thermophysical behavior of gold. Due to the extremely low electron-phonon coupling constant (G) and high electron thermal conductivity highlighted in Ref.[7], the energy absorbed by the electrons remains localized in the thin film as a result of the substrate's thermal insulation leading to the development of extreme temperature gradients. In the single-pulse irradiation regime, these gradients drive the Marangoni instabilities to an extreme point; the fluid dynamics reveals that the resulting thermocapillary effects which are significantly high for the 32 nm thick film, lead to the production of non uniform regions (i.e. with the occurrence of hot spots) observed in the SEM analysis. On the other hand, our results manifested the capacity to fabricate H-LIPSS (Figure 7a) by using double pulse irradiation and modulating the interpulse delay (Figure 7b). It was demonstrated that the 'synergy' of the EM and fluid dynamics processes for double-pulse irradiation, leads to the generation of uniform patterns. Thus, by optimizing the interpulse delay to $\Delta\tau\approx$1.2 ns (Figure 7c), we can effectively control the material's transient thermodynamic state. In other words, the technique represents a hydrodynamic guidance of the molten region via modulation of the pulse separation. An analysis of the multiscale and multiphysics processes showed that the delay allows for a partial relaxation of the electron temperature and a more gradual lattice heating, which controls efficiently the peak Marangoni stress. By maintaining the fluid velocity below the thermocapillary flow threshold, the molten gold is allowed to reorganize into the EM-prescribed periodic relief without reaching ablation conditions (i.e. which is a necessary condition for the formation of the hot spots).

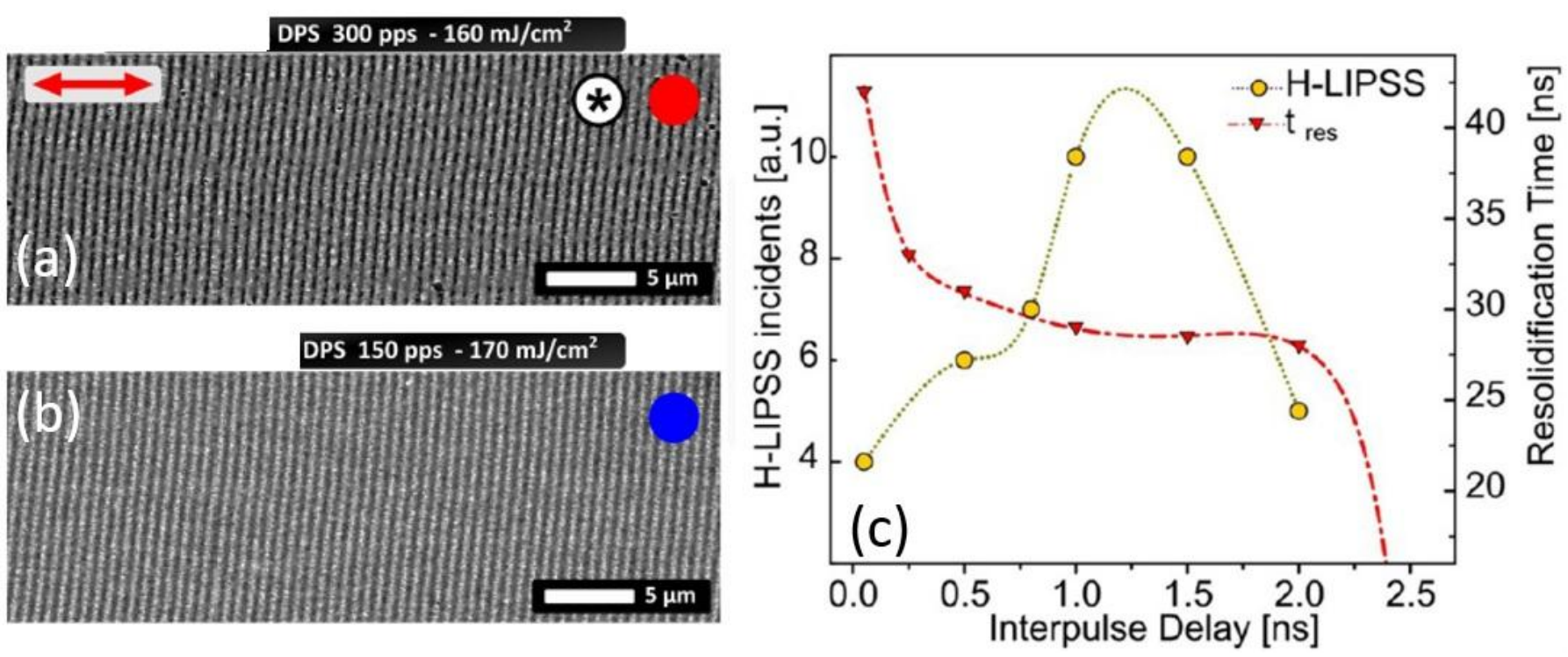

Figure 7: Highly regular LIPSS (H-LIPSS) fabricated with double pulses at (a) 300 pulses per spot (pps), 160 mJ/cm$^2$, (a) 150 pulses per spot (pps), 170 mJ/cm$^2$. (c) number of events of the production of H-LIPSS as a function of the interpulse delay.

The validation of the structural ordering achieved in this work is reflected in the optical performance of the fabricated arrays. A transition from a disordered to an ordered topography transforms the plasmonic response from broad, localized resonances to ultranarrow Surface Lattice Resonances. Spectroscopic measurements confirm that the homogeneity of the nanostructures directly governs the resonance sharpness. We demonstrated the generation of narrow diffraction-coupled SLRs and their associated phase features using periodic nanostripe arrays produced via femtosecond laser processing. Compared to conventional Electron Beam Lithography or Focused Ion Beam methods, this approach is significantly more cost-effective, scalable and capable of patterning large-area gold films without the need for high-vacuum conditions or a complex lithography. Although previous laser-assisted techniques have enabled the fabrication of plasmonic arrays and abrupt phase features, they generally relied on additional photolithography, nanoparticle patterning, or electroless plating, particularly for three-dimensional structures [18, 19]. Our method bypasses these extra steps, achieving direct laser structuring of thin metal films. This efficient, lithography-free process offers a robust pathway for large-scale production of phase-sensitive plasmonic devices with potential applications in ultrasensitive biosensing. Furthermore, understanding ultrafast fluid dynamics during laser processing is crucial to unlocking advanced photonic functionalities in these plasmonic architectures.

## 6. CONCLUSION

This study demonstrates the successful fabrication of highly-ordered LIPSS on 32-nm Au films through the synergy between coupled plasmonic modes and liquid-phase control. Our theoretical framework reveals that for thin-film geometries, the pattern periodicity is dictated by coupled SPP modes excited at both the air/metal and metal/substrate interfaces. We demonstrate that the intrinsically low electron-phonon coupling and high thermal conductivity of gold favour the development of complex Marangoni-driven hydrodynamic flows under single-pulse irradiation resulting in non-uniform surface patterns. In contrast, the implementation of a double-pulse scheme with an optimized 1.2 ns inter-pulse delay effectively suppresses these hydrodynamic instabilities leading to significantly improved pattern uniformity. By modulating the transient thermal gradients, we establish a controlled hydrodynamic regime in the molten film that inhibits thermocapillary instabilities resulting in highly ordered and ultra-regular surface arrays. These highly-ordered structures support narrow Surface Lattice Resonances, providing a high-throughput, lithography-free pathway for the development of next-generation, ultrasensitive plasmonic biosensing platforms.

## ACKNOWLEDGEMENTS

This work was supported by the EU's H2020 framework programme for research and innovation under the NFFAEurope-Pilot project (Grant No. 101007417). V.G.K. and A.N.G. acknowledge support of Graphene Flagship programme, Core 3 (881603). A.V.K., F.F., and E.S. acknowledge the contribution of the International Associated Laboratory (LIA) "Laser Matter Interaction from fundamental studies to inNOvative laSer processing- MINOS". A.V.K. acknowledges support from the French government under the France 2030 investment plan, as part of the Initiative d'Excellence d'Aix-Marseille Université_A*MIDEX" AMX-22-RE-AB-107. G.D.T. and E.S. also acknowledge funding from Horizon Europe, the European Union's Framework Programme for Research and Innovation, under Grant Agreement No. 101057457 (METAMORPHA).